# Ontological Determinism, non-locality and the system problem in quantum mechanics

**Maurice Passman**
Adaptive Risk Technology, Ltd.
London, UK
info@ariskt.co.uk

**Philip V. Fellman**
Suffolk University
Boston, MA
Shirogitsune99@yahoo.com

**Jonathan Vos Post**
Computer Futures
Altadena, CA
Jvospost3@gmail.com

# 1 Introduction

In "The Trouble With Physics", Lee Smolin recounts the philosophical dilemmas of modern physical systems theory. [1] It is easy enough to paraphrase Smolin when illustrating the breakdown of classical physical systems concepts and the incomplete picture of knowledge (indeed the necessarily incomplete picture for those who know their Gödel). Any observer of modern physics can see there are several areas where the theory must either cohere or perhaps be rejected and replaced (in the long run if not the short.) This is not to argue in favor of "crackpot" radical schemes or to suddenly reject the advances which general relativity and quantum mechanics have made in resolving problems which classical methods failed to solve. Nonetheless, physics is in deep philosophical and theoretical difficulty today, especially in its more esoteric branches. This is rather evident if one simply skates across Stuart Kauffman's "boundary between order and chaos" [2] to survey where physical theory begins to break down.

For example, Newtonian kinematics runs into trouble with three bodies as Poincaré explained in pioneering what became chaos theory, even though for two bodies we get lovely Keplerian dynamics (as Feynman gave in his original elementary exposition).[3] General Relativity runs into trouble with one body at two scales: (a) Very small bodies, length near or less than a Planck length, and totally breaking down as length approaches zero; as well as (b) the universe(s) as a whole. Likewise, Quantum Field Theory also has trouble with one body, as it interacts with its own field, in a way which requires renormalization. String Theory runs into trouble with zero bodies, as it predicts $10^{500}$ or $10^{1000}$ different vacuums, of which we don't know which one we ever had.

In this paper we will review the "measurement" problem in quantum mechanics, following our earlier paper, "The Fundamental Importance of Discourse in Theoretical Physics" [4], and paying especial attention to the arguments of John S. Bell, particularly those dealing with Bohr's "debate" with Einstein on the language and meaning of quantum mechanics. From there we will enter into a more technical treatment of non-locality as well as demonstrating that even where quantum mechanics is deterministic, this is not an ontological necessity.

## 1.1 Quantum Mechanics – Interpretations and Non-locality

John Bell gives a wonderful summary of the role of discourse in theoretical physics and the difficulties of coming to grips with it in his 1984 paper, "Bertlmann's Socks and the Nature of Reality"[5], beginning with Einstein's argument that:

> If one asks what, irrespective of quantum mechanics, is characteristic of the world of ideas of physics, one is first of all struck by the following: the concepts of physics relate to a real outside world…It is further characteristic of these physical objects that they are thought of as arranged in a space-time continuum. An essential aspect of this arrangement of things in physics is that they lay claim, at a certain time, to an existence independent of one another, provided these objects 'are situated in different parts of space.

This is essentially the philosophical position advocated by Ludwig Wittgenstein in the "Tractatus Logico-Philosophicus"[6]. Even without the benefit of subsequent arguments, including those by Wittgenstein himself, who ultimately rejected this viewpoint, [7]. Bell gives us Einstein's own qualification from quantum mechanics:

> "There seems to me no doubt that those physicists who regard the descriptive method of quantum mechanics as definitive in principle would react to this line of thought in the following way: they would drop the requirement…for the independent existence of the physical reality present in different parts of space; they would be justified in pointing out that the quantum theory nowhere makes explicit use of this requirement."

Einstein then goes on to complete the argument by explaining that he nonetheless sees no reason why this requirement would necessarily have to be abandoned and that this was why he believed quantum mechanics to be an incomplete theory. What is noteworthy in regard to Bell's raising this argument is that the contemporary discourse of the opposing school of thought is even less satisfying than the explanation provided by Einstein. Bell's argument begins with the problematical position of Bohr. "Bohr once declared when asked whether the quantum mechanical algorithm could be considered as somehow mirroring an underlying quantum reality: 'There is no quantum world. There is only an abstract quantum mechanical description. It is wrong to think that the task of physics is to find out how Nature is. Physics concerns what we can say about nature." Not only is this bad philosophical discourse, but it is bad physics. Bell illustrates the scope of Bohr's problem by noting that:

> While imagining that I understand the position of Einstein as regards EPR correlations, I have very little understanding of his principal opponent, Bohr. Yet most contemporary theorists have the impression that Bohr got the better of Einstein in the argument and are under the impression that they themselves share Bohr's views. As an indication of those views, I quote a passage from his reply to Einstein, Podolsky and Rosen. It is a passage which Bohr himself seems to have regarded as definitive, quoting it himself when summing up much later. Einstein, Podolsky and Rosen had assumed that '…if, without in any way disturbing a system, we can predict with certainty the value of a physical quantity, then there exists an element of physical reality corresponding to this physical quantity'. Bohr replied: '…the wording of the above mentioned criterion…contains an ambiguity as regards the meaning of the expression 'without in any way disturbing a system'. Of course there is in a case like that just considered no question of mechanical disturbance of the system under investigation during the last critical stages of the measuring procedure. But even at this stage there is essentially the question of *an influence on the very conditions which define the possible types of predictions regarding the future behavior of the system*…their argumentation does not justify their conclusion that quantum mechanical description is essentially incomplete…this description may be characterized as the rational utilization of all possibilities of unambiguous interpretation of measurements, compatible with the finite and uncontrollable action between the objects and the measuring instruments in the field of quantum theory'

Finally, Bell explains why this is both bad philosophy and bad physics:

> Indeed, I have very little idea what this means. I do not understand in what sense the word 'mechanical' is used, in characterizing the disturbances which Bohr does not contemplate, as distinct from those which he does. I do not know what the italicized passage means – "an influence on the very conditions…" Could it mean just that different experiments on the first system give different kinds of information about the second? But this was just one of the main points of EPR, who observed that one could learn *either* the position *or* the momentum of the second system. And then I do not understand the final reference to 'uncontrollable interactions between measuring instruments and objects.' It seems just to ignore the essential point of EPR that in the absence of action at a distance, only the first system could be supposedly disturbed by the first measurement and yet definite predictions become possible for the second system. Is Bohr just rejecting the premise - "no action at a distance" – rather than refuting the argument?

Indeed, it does appear that Bohr simply rejects non-locality, and as our subsequent analysis shows, this is both mathematically and empirically bad science. While we do not pretend to have the complete philosophical position necessary to accommodate all the various aspects of non-locality, we do have some clear notions of what physics tells us about non-locality, and how we can understand the foundations of non-local interactions without resorting to such cumbersome concepts as "non-reality".

## 2.1 Non-Locality

Here we recapitulate our arguments on non-locality from "Time, Uncertainty and Non-Locality in Quantum Cosmology" [8]. We begin with the Schrödinger wave equation and continue through an exposition of non-locality in detail, covering quantum entanglement and the Stern-Gerlach apparatus (which for the purposes of our discussion directly addresses the issue of the "quantum system" and the quantum problem of measurement), including a brief discussion of quantum mechanical determinism and ontological necessity.

## 3.1 The Schrodinger Wave Equation

Wave functions live on configuration space. Schrödinger called this entanglement. The linearity of the Schrödinger equation prevents the wave function from representing reality. If the equation were non-linear (e.g. reduction models) the wave function living on configuration space still could not by itself represent reality in physical space.[5] Given a system that can be described by linear combinations of wave functions $\varphi_1$ and $\varphi_2$. We also have a piece of apparatus that, when brought into interaction with the system, measures whether the system has wave function $\varphi_1$ and $\varphi_2$. Measuring means that, next to the 0 pointer position, the apparatus has two pointer positions, 1 and 2, 'described' by the wave functions $\Psi_0, \Psi_1, \Psi_2$ for which

$$\varphi_i \Psi_0 \xrightarrow{Schrodinger\ evolution} \varphi_i \Psi_i, \quad i = 1, 2$$

The wave function has a support in configuration space which corresponds classically to a set of coordinates of particles (which would form a pointer). For superposition:

$$\varphi = c_1 \varphi_1 + c_2 \varphi_2, \quad c_1, c_2 \in \mathbb{C}, \quad |c_1|^2 + |c_2|^2 = 1$$

$$\varphi_i \Psi_0 = (c_1 \varphi_1 + c_2 \varphi_2) \xrightarrow{Schrodinger\ evolution} c_1 \varphi_1 \Psi_1 + c_2 \varphi_2 \Psi_2$$

The outcome on the right side does not concur with experience. It shows a 'macroscopic indeterminacy'. For the Schrödinger cat experiment $\varphi_1$ and $\varphi_2$ are the wave functions of the non-decayed and the decayed atom; $\Psi_0$ and $\Psi_1$ are the wave functions for the live cat and $\Psi_2$ is the wave function for the dead cat. Schrödinger says that this is unacceptable. Why? Isn't the apparatus supposed to be the observer? What counts as an observer?

The evolution of $\varphi_i \Psi_0 = (c_1 \varphi_1 + c_2 \varphi_2) \xrightarrow{Schrodinger\ evolution} c_1 \varphi_1 \Psi_1 + c_2 \varphi_2 \Psi_2$ is an instance of decoherence. The apparatus decoheres the superposition $(c_1 \varphi_1 + c_2 \varphi_2)$ of the system wave function. Decoherence means that it is in a practical sense impossible to get the two wave packets $\varphi_1 \Psi_1$ and $\varphi_1 \Psi_2$ superimposed in $c_1 \varphi_1 \Psi_1 + c_2 \varphi_2 \Psi_2$ to interfere. Decomposition is this practical impossibility – Bell referred this as 'fapp-impossibility' where fapp = for all practical purposes.[7]

## 3.2 Mechanics and the Double-Slit Experiment

Particle motion is guided by the wave function. The physical theory is formulated with the variables $\mathbf{q}_i \in \mathbb{R}^3, \quad i = 1, 2, 3, ...N$, the positions of the $N$ particles that make up the system, and the wave function $\psi(\mathbf{q}_1, ....\mathbf{q}_N)$ on the configuration space of the system. Quantum randomness – Born's statistical law – is explained on the basis of Bolzmann's principles of statistical mechanics. Born's law is not an axiom but a theorem; Born's statistical law concerning $\rho = |\psi|^2$ is that if the wave function is $\psi$ then the particle configuration is $|\psi|^2$-distributed. Applying this to $\varphi_i \Psi_0 = (c_1 \varphi_1 + c_2 \varphi_2) \xrightarrow{Schrodinger\ evolution} c_1 \varphi_1 \Psi_1 + c_2 \varphi_2 \Psi_2$ above implies that the result $i$ comes with probability $|c_i|^2$.

In the double slit experiment, each particle goes either through the upper or through the lower slit. The wave function goes through both slits and forms after the slits a wave function with an interference pattern:

(A) Close slit 1 open slit 2; Particle goes through slit 2; It arrives at on the plate with probability $|\psi_2(\mathbf{x})|^2$

Where $\psi_2$ is the wave function which passed though slit 2.

(B) Close slit 2 open slit 1; Particle goes through slit 1. It arrives at **x** on the plate with probability $|\psi_1(\mathbf{x})|^2$

Where $\psi_1$ is the wave function which passed though slit 1.

(C) Both slits are open; The particle goes through slit 1 or slit 2. It arrives at **x** on the plate with probability $|\psi_1(\mathbf{x})|^2 + |\psi_2(\mathbf{x})|^2$

In general:

$$|\psi_1(\mathbf{x}) + \psi_2(\mathbf{x})|^2 = |\psi_1(\mathbf{x})|^2 + |\psi_2(\mathbf{x})|^2 + 2\Re\,\psi_1^*(\mathbf{x})\psi_2(\mathbf{x}) \neq |\psi_1(\mathbf{x})|^2 + |\psi_2(\mathbf{x})|^2$$

Here $\Re$ denotes the real part of a complex quantity. The inequality comes from the interference of the wave packets $\psi_1$, $\psi_2$ which passes through slit 1 and 2. Situations 'Particle goes through slit 2' and 'Particle goes through slit 1' are exclusive alternatives entering 'The particle goes through slit 1 or slit 2', but the probabilities $|\psi_2(\mathbf{x})|^2$ and $|\psi_1(\mathbf{x})|^2$ do not add up – this is because 'Close slit 1 open slit 2', Close slit 1 open slit 1' and 'Both slits are open' are *physically distinct*.

## 4.1 Causality, Determinism and Ontology

This type of methodology is often said to aim at restoring determinism to the quantum world. Determinism has nothing to do with ontology. This type of QM is deterministic - but is not an ontological necessity. In this regard, our position (derived from relatively straightforward Bohmian mechanics) is rather different than that of T'Hooft [14], particularly in "Quantum Mechanics and Determinism', where the effort is directed at mapping the quantum states of a system of free scalar particles one-to-one onto the states of a completely deterministic model. T'Hooft has several schema of this type, including mappings applied to free Maxwell fields and demonstrating Lorentz invariance. [15][16] This is not to deny that T'Hooft has demonstrated some brilliant insights into the epistemological nature of quantum mechanics and quantum field theory. He certainly argues with great plausibility that string theory cannot fulfill the role of the standard model for quantum field theory.[1] [16]

---

[1] Specifically, "For the last couple of decades, theoreticians have been considering the challenge to include the one remaining theoretical requirement: general relativity, i.e., the laws of gravity. The problem appears to be a beautiful one, since now the dynamics of the curvature in space and time must be submitted to the laws of quantum mechanics. The virtual contributions to the amplitudes due to space-time curvature diverge at the small-distance end, just like Fermi's earliest models for the weak force used to do.

However, by his own admission, while he finds merit in approaching dynamics below the Planck scale as being so complex that they lead to apparently stochastic fluctuations (a phenomenon well known in non-linear dynamical systems modeling) that he models the system with a first order perturbative theory, even though this type of theory leads to the EPR paradox and Bell's inequalities. Part of our difficulty with T'Hooft's approach is that in this kind of method, "…the definition of time does not need to be very strict. One might have a continuous time variable or discretized time, or time might be defined in terms of Cauchy surfaces in a general relativistic setting. There is one important condition that must be met by the time variable, however: it must be on a real line (possibly with a beginning and an end to it). Time is not allowed to be cyclic. If closed time-like trajectories would exist, this would lead to clashes and our theory would no longer be unambiguous. Closed time-like loops, popular in some versions of gravity theories, will no longer be excluded."[16]

However it is precisely the role of time which has been the focus of our investigations in quantum cosmology. [8][17][18][19][20][21][22]. In this regard, the time function is critically important and our interpretation of quantum mechanics, and indeed quantum cosmology necessarily precludes a flow of static infinitesimals, prohibits discretization and does not prohibit global cyclicality [22]. Although we have not fully explored the mathematical and topological landscape of closed timelike geodesics, there is at present, nothing in our proposed theory which categorically rules out the existence of closed time-like trajectories. [18] Additionally, T'Hooft must introduce a number of complex constraints which limit the observables of his theory, particularly when he must deal with non-locality.[14] In contradistinction, we are able to relate observables and even the characterization of the observational apparatus directly to non-locality.

## 4.2 Locality

As we argued in a previous exposition [8], Einstein deduced from Maxwell's equations that space and time change in a different way from Galilean physics when one changes between frames moving with respect to each other. The nature of this change is governed by the unchanging velocity of light when moving from one frame to another. This led to Minkowski showing that a particle needs a position in time and space for its specification therefore implying that a particle in relativistic space should have time and space coordinates.

In this section, we explore the argument that any theory must be nonlocal and attempt to present a mathematical proof to that effect. Nonlocality is crudely defined as meaning that the theory contains action at a distance in the true meaning of the words i.e. faster than light action between separated events.

Action at a distance is such that no information can be sent with superluminal speed – therefore, there is no inconsistency with special relativity. Nonlocality is encoded in the wave function that lives on configuration space and is by its very nature a nonlocal agent. All particles are guided simultaneously by the wave function

---

In the latter case, the divergences were successfully tamed by the introduction of the Standard Model. Naturally, one expects similar solutions to the problem at hand, and indeed, string theories and their successors are claimed to come close to bringing just such a solution. But, they have not done so yet, and upon closer inspection one finds that there really are reasons to be skeptical. Not only are there numerous technical difficulties — the identification of the ground state, our lack of understanding the supersymmetry breaking mechanism, the smallness of the cosmological constant —,there are also more fundamental ones, which are difficulties that have little to do with the fact that we are dealing with strings, D-branes, or what not. Rather, they have to do with the fact that we are attempting to apply quantum mechanics to dynamical laws that should be the ultimate driving forces of the entire Universe; one will be forced to consider statistical ensembles of universes, and such notions will be much more questionable than the notion of an ensemble of experiments inside one universe. It should be kept in mind that one will never be able to do experiments in more than one universe, and that the 'averaged value' of a quantity measured in different universes cannot be checked against any theory.

and *f* the wave function is entangled, the nonlocal action does not get small with particles.

In a two particle system with coordinates $\mathbf{X}_1(t), \mathbf{X}_2(t)$ we have:

$$\dot{\mathbf{X}}_1(t) = \frac{\hbar}{m_1} \Im \left[ \frac{\left.\frac{\partial}{\partial \mathbf{x}} \psi(\mathbf{x}, \mathbf{X}_2(t))\right|_{\mathbf{x}=\mathbf{X}_1(t)}}{\psi(\mathbf{X}_1(t), \mathbf{X}_2(t))} \right]$$

Therefore the velocity of $\mathbf{X}_1(t)$ at time *t* depends in general on $\mathbf{X}_2(t)$ at time t, no matter how far apart the positions are. In general here means that the wave function is entangled and not a product e.g.: $\psi(\mathbf{x},\mathbf{y}) = \varphi(\mathbf{x})\Phi(\mathbf{y})$. However, there is no immediate reason why the wave function should become a product when **x** and **y** are far apart (although decoherence is always lurking awaiting an opportunity to destroy coherence i.e. produce an effective product structure). This exposition in expanded upon in Appendix I, "Bell's Theorem".

## 5.1 Statistical Ontology, Non-Locality and "The Apparatus"

Justification for Born's statistical interpretation of the wave function is based upon typicality arguments. This gives a hypothesis for quantum equilibrium. If a subsystem has effective wave function $\varphi$ then its particle coordinates are $|\varphi|^2$ distributed. Proof of this hypothesis entails a proof of the law of large numbers. If we return to the Stern-Gerlach magnets as in our previous examples, from our previous references [21][22][23][24][25] and Dürr, Goldstein and Zanghì (1992)[27], we need to demonstrate that there is no effect on the statistics of outcomes; they are the same whether we take a measurement on the magnet on the left, or a measurement on the magnet on the right, first.

Take a general entangled two-particle state:

$$\psi = a|\uparrow\rangle_1|\downarrow\rangle_2 + b|\downarrow\rangle_1|\uparrow\rangle_2 + c|\downarrow\rangle_1|\downarrow\rangle_2 + d|\uparrow\rangle_1|\uparrow\rangle_2$$

$$\text{with } |a|^2 + |b|^2 + |c|^2 + |d|^2 = 1.$$

The probability of getting the spin value $|\uparrow\rangle_2$ at the right magnet is $|b|^2 + |d|^2$. Now undertake a measurement at the left magnet in an arbitrary direction $\gamma$, where $\gamma$ is the angle between the chosen direction and the z-direction. We can now write the entangled equation in the $\gamma$-basis:

$$|\uparrow\rangle_1 = \mathbf{i}_1 \cos\gamma + \mathbf{j}_i \sin\gamma \text{ and } |\downarrow\rangle_1 = -\mathbf{i}_1 \sin\gamma + \mathbf{j}_i \cos\gamma$$

We can rewrite the state $\psi$ as:

$$\psi = \mathbf{i}_1\left[\cos\gamma(a|\downarrow\rangle_2 + d|\uparrow\rangle_2) - \sin\gamma(b|\uparrow\rangle_2 + c|\downarrow\rangle_2)\right] + \mathbf{j}_1\left[\sin\gamma(a|\downarrow\rangle_2 + d|\uparrow\rangle_2) + \cos\gamma(b|\uparrow\rangle_2 + c|\downarrow\rangle_2)\right]$$

$$\approx \psi_{\mathbf{i}_1} + \psi_{\mathbf{j}_1}$$

Therefore $\|\psi_{\mathbf{i}_1}\|^2$ and $\|\psi_{\mathbf{j}_1}\|^2$ are the probabilities for the outcomes spin up or spin down when measuring first at the left magnet at an arbitrary angle $\gamma$. Therefore the collapsed wave function will either be $\dfrac{\psi_{\mathbf{i}_1}}{\|\psi_{\mathbf{i}_1}\|}$ or $\dfrac{\psi_{\mathbf{j}_1}}{\|\psi_{\mathbf{j}_1}\|}$ depending upon the outcome at the left hand magnet.

The effect of this measurement on the probability for the outcome at the right magnet is as follows.

Calculating the equilibrium probability for the spin up wave function $|\uparrow\rangle_2$, for the collapsed wave function $\dfrac{\psi_{\mathbf{i}_1}}{\|\psi_{\mathbf{i}_1}\|}$, we obtain the probability:

$$\frac{\left\|\mathbf{i}_1 |\uparrow\rangle_2 (d\cos\gamma - b\sin\gamma)\right\|^2}{\left\|\psi_{\mathbf{i}_1}\right\|^2}$$

And for $\dfrac{\psi_{\mathbf{j}_1}}{\left\|\psi_{\mathbf{j}_1}\right\|}$, we obtain:

$$\frac{\left\|\mathbf{j}_1 |\uparrow\rangle_2 (b\cos\gamma - d\sin\gamma)\right\|^2}{\left\|\psi_{\mathbf{j}_1}\right\|^2}$$

The outcome for the spin up case for the right hand magnet is therefore:

$$\left\|\psi_{\mathbf{i}_1}\right\|^2 \frac{\left\|\mathbf{i}_1 |\uparrow\rangle_2 (d\cos\gamma - b\sin\gamma)\right\|^2}{\left\|\psi_{\mathbf{i}_1}\right\|^2} + \left\|\psi_{\mathbf{j}_1}\right\|^2 \frac{\left\|\mathbf{j}_1 |\uparrow\rangle_2 (b\cos\gamma - d\sin\gamma)\right\|^2}{\left\|\psi_{\mathbf{j}_1}\right\|^2} = |b|^2 + |d|^2$$

Therefore the 'statistical' outcomes are the same no matter if the measurement takes place on the left or right hand magnet.

## 6  Conclusion

The role and nature of determinism in quantum mechanics is complex, and is hardly a closed question. On the one hand, T'Hooft has contributed important elements of an emerging framework for deterministic quantum mechanics (within the framework of quantum-mechanical probability distributions, or what F.S.C. Northrop refers to as "theoretical probability"). On the other hand, the ability of a pseudo-classical stochastic perturbative process to accurately model Planck scale behavior in a fine-grained fashion has yet to be established. More comprehensively, outside of invoking decoherence, non-locality appears to be key component of any quantum mechanics, much as many authors would like to find a way around it. Time too has a role in this model, not as an independent physical variable, so much as the preserver of continuity in nature. Different approaches to time yield very different models of quantum mechanical behavior and appear to have a fundamental role in quantum cosmology.

# Appendix I: Bell's Theorem[2]

Is it possible to describe a quantum mechanics that is local? Einstein, Podoloski and Rosen (EPR) thought so. The EPR argument is interesting as it constitutes part of Bell's proof of the Nonlocality of nature. Bell's response is Bell's theorem (i.e., nature is nonlocal).

In this first part of the argument on Bell's Theorem, the EPR argument applied to a simplified EPR Gedanken experiment. This approach is based upon the argument that one can prepare a special pair (L, R) of spin ½ particles that fly apart in opposite directions and which behave in the following well determined fashion.

When both particles pass identically oriented Stern-Gerlach magnets, they deflect in exactly opposite direction. If L has **a**-spin +1/2 then R has **a**-spin -1/2 where **a** is the orientation of the magnets. This is true for all directions **a**. The probability for L up, R down is ½. The two particle wave function is called a singlet state and the total spin of this singlet state is zero. Measuring first the **a**-spin on L, we can predict with certainty the result of the measurement of the **a**-spin on R. This is true even of the measurement events L and R are space-like separated.

Suppose that the experiment is arranged in such a way that a light signal cannot communicate the L result to the R particle before the R particle passes SGM-R. Suppose now that 'locality' holds meaning that the spin measurement on one side has no superluminal influence on the result of the spin measurement on the other side. Then we must conclude that the value we predict for the **a**-spin on R is pre-existing. It cannot have been created by the result obtained on L as we assume locality. If the value pre-exists, then that means that it exists even before the decision was taken in which direction **a** the spin on the left is to be measured. The value therefore pre-exists for any direction **a**. This also holds (by symmetry) for the values obtained on L. by locality, therefore we obtain the pre-existing values of spins on either side in any direction.

We collect the pre-existing values in a family of variables $\mathbf{X}_\mathbf{a}^{(L)}, \mathbf{X}_\mathbf{a}^{(R)} \in \{-1,1\}$ with **a** indexing arbitrary directions and with $\mathbf{X}_\mathbf{a}^{(L)} = \mathbf{X}_\mathbf{a}^{(R)}$. The locality check is now to ask the question if such pre-existing values actually exist. However, there is no way that the variables can reproduce the quantum mechanical correlations. Choose three directions, given by unit vectors **a**, **b**, and **c** and consider the corresponding 6 variables:

$\mathbf{X}_\mathbf{y}^{(L)}, \mathbf{X}_\mathbf{z}^{(R)}, \mathbf{y}, \mathbf{z} \in \{\mathbf{a},\mathbf{b},\mathbf{c}\}$. They must satisfy:

$$\left(\mathbf{X}_\mathbf{a}^{(L)}, \mathbf{X}_\mathbf{b}^{(L)}, \mathbf{X}_\mathbf{c}^{(L)}\right) = \left(-\mathbf{X}_\mathbf{a}^{(R)}, -\mathbf{X}_\mathbf{b}^{(R)}, -\mathbf{X}_\mathbf{c}^{(R)}\right)$$

We wish to reproduce the relative frequencies of the anticorrelation events:

$$\mathbf{X}_\mathbf{a}^{(L)} = -\mathbf{X}_\mathbf{b}^{(R)}, \quad \mathbf{X}_\mathbf{b}^{(L)} = -\mathbf{X}_\mathbf{c}^{(R)}, \quad \mathbf{X}_\mathbf{c}^{(L)} = -\mathbf{X}_\mathbf{a}^{(R)}$$

Adding the probabilities and using the rules of probability, we get:

$$\text{Prob}\left(\mathbf{X}_\mathbf{a}^{(L)} = -\mathbf{X}_\mathbf{b}^{(R)}\right) \;+\; \text{Prob}\left(\mathbf{X}_\mathbf{b}^{(L)} = -\mathbf{X}_\mathbf{c}^{(R)}\right) \;+\; \text{Prob}\left(\mathbf{X}_\mathbf{c}^{(L)} = -\mathbf{X}_\mathbf{a}^{(R)}\right)$$

---

[2] Abstracted from Passman, Maurice, Fellman, Philip V., and Post, Jonathan Vos (2011) Time, Uncertainty and Non-Locality in Quantum Cosmology, paper presented at the 8[th] International Conference on Complex Systems, Quincy, MA June 26-July 1, 2011.

$$= \text{Prob}\left(X_a^{(L)} = X_b^{(L)}\right) + \text{Prob}\left(X_b^{(L)} = X_c^{(L)}\right) + \text{Prob}\left(X_c^{(L)} = X_a^{(L)}\right)$$

$$\geq \text{Prob}\left(X_a^{(L)} = X_b^{(L)} \quad \text{or} \quad X_b^{(L)} = X_c^{(L)} \text{or} \quad X_c^{(L)} = X_a^{(L)}\right)$$

= Prob (sure event) = 1, as $\mathbf{X}_y^{(i)}, i = L, R, \mathbf{y} \in \{\mathbf{a}, \mathbf{b}, \mathbf{c}\}$ can only take two values.

This is therefore one version of Bell's inequality:

$$\text{Prob}\left(\mathbf{X}_a^{(L)} = -\mathbf{X}_b^{(R)}\right) + \text{Prob}\left(\mathbf{X}_b^{(L)} = -\mathbf{X}_c^{(R)}\right) + \text{Prob}\left(\mathbf{X}_c^{(L)} = -\mathbf{X}_a^{(R)}\right) \geq 1$$

The logical structure of Bell's nonlocality argument follows.

Let P be the hypothesis of the existence of pre-existing values

$X_{a,b,c}^{L,R}$ for the spin components relevant to this EPRB experiment. Then:

First part: quantum mechanics + locality -> P

Second part: quantum mechanics -> not P

Conclusion: quantum mechanics -> nonlocality

$$\text{Prob}\left(\mathbf{X}_a^{(L)} = -\mathbf{X}_b^{(R)}\right) + \text{Prob}\left(\mathbf{X}_b^{(L)} = -\mathbf{X}_c^{(R)}\right) + \text{Prob}\left(\mathbf{X}_c^{(L)} = -\mathbf{X}_a^{(R)}\right) \geq 1$$

is violated by experimental evidence (Aspect et Al, 1982a, 1982b). [25][26]

We could therefore write:
Bell's Theorem - First part: experimental facts + locality -> P
Bell's Theorem - Second part: experimental facts -> not P
Conclusion: experimental facts -> nonlocality
Nature is therefore nonlocal. QED!